\documentclass[twocolumn,amssymb,amsmath,aps,prb,showpacs,floatfix]{revtex4-2}
\usepackage{array}
\usepackage{graphicx}
\usepackage{trimclip}
\usepackage{accents}
\usepackage{color}
\usepackage{multirow}
\usepackage{float}
\usepackage{amssymb}
\usepackage{gensymb}
\usepackage{amsmath}
\usepackage{ulem}
\usepackage{makecell}
\usepackage[urlcolor=blue,colorlinks=true,citecolor=blue,linkcolor=blue,pdfstartview={FitH },bookmarks=false]{hyperref}
\usepackage[utf8]{inputenc}
\graphicspath{{fig/}{./fig/}{.}}

\usepackage{scrextend}
\usepackage{orcidlink}

\begin{document}
	\title{Thermodynamic and electrical transport properties of the half-Heusler plumbide TbAuPb}
	\author{Abhinav Agarwal\orcidlink{0009-0006-6332-9374}}
	\author{Snehashish Chatterjee \orcidlink{0000-0002-2324-0096}}
 \author{Maciej J. Winiarski \orcidlink{0000-0001-6935-4825}}
 \author{Orest Pavlosiuk \orcidlink{0000-0001-5210-2664}}
 \author{Dorota A. Kowalska \orcidlink{0000-0001-8180-653X}}
	\author{Piotr Wiśniewski\orcidlink{0000-0002-6741-2793}} 
	\author{Dariusz Kaczorowski \orcidlink{0000-0002-8513-7422}}
	\affiliation{Institute of Low Temperature and Structure Research, Polish Academy of Sciences, Okólna 2, Wroc\l{}aw 50-422, Poland}
	
	\begin{abstract}
	   Structural, thermodynamic and electrical transport properties of TbAuPb were investigated on single crystals. The compound was found to crystallize with the cubic MgAgAs-type structure characteristic of half-Heusler materials. It orders antiferromagnetically at $T_\mathrm{N}$ = 5~K and undergoes a transition into a different antiferromagnetic phase emerging in high magnetic fields. Electrical transport in TbAuPb exhibits a multiband character, with a predominance of hole-like carriers. Angular magnetoresistance evolves systematically with applied magnetic field and changes its symmetry near the spin-reorientation transition, highlighting strong coupling between the charge transport and the magnetic order. The results of first-principles calculations indicate that TbAuPb is a band inverted semimetal in the non-magnetic state, which becomes topologically trivial in the field-induced ferromagnetic state.
	\end{abstract}
	\maketitle

\section{Introduction}
The search for new quantum materials with non-trivial electronic band structures remains a central focus in experimental condensed matter physics, driven by both fundamental interest and potential technological applications \cite{Hasan2010}. Rare-earth-based half-Heusler (HH) phases $RTX$ ($R$-rare earth, $T$-transition element, and $X$-main group element) have been extensively investigated over the past decade, particularly following predictions of their potential as topological insulators or topological semimetals \cite{Al-Sawai2010, Xiao2010, Feng2010}. Moreover, HH compounds have shown a range of intriguing phenomena including heavy fermion behavior \cite{Canfield1991, Mun2013, Mun2015, Guo2018}, thermoelectric effects\cite{Wang2023, Pavlosiuk2025}, and non-trivial superconductivity \cite{Tafti2013, Nakajima2015, Meinert2016, Ishihara2021}.  Collectively, these properties make HH compounds a versatile platform, wherein the physical properties can be systematically tuned by chemical substitution \cite{Graf2011}.

The widely studied $RTX$ compounds with non-trivial band topology are the bismuthides $R$PtBi \cite{Chen2020}, which have been identified as promising candidates for Weyl semimetals \cite{Shekhar2018}, showing a plethora of intriguing physical phenomena such as anomalous Hall effect \cite{Suzuki2016, Zhu2020}, negative longitudinal magnetoresistance \cite{Hirschberger2016}, chiral magnetic anomaly \cite{Liang2018}, and planar Hall effect \cite{Kumar2018}. Recently, the research has been extended to other series of HH compounds, e.g. the stannides $R$AuSn, which were found to undergo a transformation from topologically trivial semimetal state in zero magnetic field to Weyl semimetal phase in applied magnetic field \cite{Ueda2023, Ueda2025, Lu2025, ErAuSn}. These findings highlight the importance of studying unexplored $RTX$ compounds to gain deeper insight into the interplay between magnetism and band topology, both in zero and finite magnetic fields.

The electronic band structure of the HH compounds is similar to that of zinc-blende (Hg,Cd)Te and highly tunable \cite{Chadov2010}. Non-trivial band topology arises from the inversion of $\Gamma_6$ and $\Gamma_8$ electronic states \cite{Yan2014, Chadov2010}. In topologically trivial materials, the $\Gamma_8$ states lie below the $\Gamma_6$ states. Conversely, in topologically non-trivial materials $\Gamma_8$ states are pushed above the $\Gamma_6$ states. The band inversion strength is proportional to the parameter $t=(Z_T+Z_X)V$, where $Z_T$ and $Z_X$ are the atomic numbers of $T$ and $X$ atoms and $V$ is the unit cell volume \cite{Al-Sawai2010}. Therefore, it increases with heavier constituent atoms, making band inversion more pronounced in systems containing high $Z$-elements. 

Theoretical calculations of the electronic structures of the non-magnetic HH plumbides LaAuPb, YAuPb, and LuAuPb revealed the effect of band inversion, suggesting their non-trivial topological character \cite{Al-Sawai2010}. It is therefore interesting to examine how magnetism influences such non-trivial electronic properties. In rare-earth-based materials, the 4\textit{f} orbitals are well shielded by 5\textit{s}, 5\textit{p} and 4\textit{d} orbitals and thus weakly participate in chemical bonding. Nevertheless, they give rise to localized magnetic moments that order at low-temperatures.  Magnetic ordering can substantially modify the electronic structure through Zeeman and/or exchange splitting, as demonstrated in several systems that are topologically trivial in the non-magnetic state but become a Weyl semimetal due to the crossing of the spin-split bands \cite{Suzuki2016, Shekhar2018, Soh2019, Chen2021}. Most recently, Liu et al. reported GdAuPb to be a topological semimetal, which exhibits a large anomalous Hall conductivity, a positive unsaturated magnetoresistance, and easily observed quantum oscillations \cite{Liu2024}. These findings motivated us to undertake a systematic study of the $R$AuPb series, for which experimental data remains scarce. In particular, the magnetic and transport properties of a novel compound TbAuPb have not been previously investigated. This  compound has two important features. First, all its constituent elements are heavy, resulting in $t$ = 49.4 $\mathrm{nm^{3}}$ which is nearly the same as those of GdAuPb ($t$ = 50 $\mathrm{nm^{3}}$) and GdPtBi ($t$ = 48 $\mathrm{nm^{3}}$), another well-established band-inverted semimetal \cite{Chadov2010, Suzuki2016, Shekhar2018}. Second, the presence of partially filled 4\textit{f} orbitals introduces intrinsic magnetism into the system. It is also noteworthy that unlike Gd, for which the orbital angular momentum is zero and crystal electric field (CEF) is negligible, Tb carries an orbital angular momentum, resulting in strong CEF interactions that can further influence the magnetic and transport properties. These considerations highlight the importance of  synthesizing and characterizing new single crystals, which may further broaden our understanding of different HH phases.

In this work, the thermodynamic  and electrical transport properties (electrical resistivity, magnetoresistance, angular magnetoresistance and Hall resistivity) of this material were studied on single crystals over wide temperature and magnetic field ranges. The experimental observations were confronted with the results of our electronic band structure calculations from first principles. 
    
\section{Methods}
\subsection{Experimental details}

Single crystals of TbAuPb were prepared using the self-flux method. High-purity constituent elements were taken in the molar ratio 1:1:50, placed in an alumina crucible, and sealed in an evacuated quartz tube. The ampule was heated in a vertical resistance furnace to 1050\degree C over 10 hours, kept at this temperature for 24 hours, and then cooled down to 500\degree C with a rate of 4\degree C per hour. Subsequently, the ampule was taken out of the furnace and centrifuged to remove Pb flux. The process yielded triangular-shaped crystals with typical dimensions 1$\times$ 1$\times$ 1 $\mathrm{mm^{3}}$. The observed morphology indicates that the crystals naturally grew along the [111] crystallographic axis. 
    
Chemical composition of the obtained crystals was examined using a FEI scanning electron microscope (SEM) equipped with a Genesis XM4 energy dispersive X-ray (EDX) probe. The results proved their single-phase nature with almost equiatomic stoichiometry (see Supplementary Material). The crystal structure of TbAuPb was verified by single-crystal X-ray diffraction carried out on an Oxford Diffraction Xcalibur four-circle diffractometer (Mo-K$\alpha$ radiation) with a CCD camera. The compound crystallizes with a non-centrosymmetric MgAgAs-type structure (space group $F\bar{4}3m$) inherent to the HH phases (see Supplementary Material). The experimental lattice parameter $a$ = 6.745~\text{\AA} is somewhat smaller than that reported for GdAuPb ($a$ = 6.774~\text{\AA} \cite{Liu2024}), in concert with the regular lanthanide contraction effect. 
    
Magnetic measurements were performed in the temperature range 2-300~K and in magnetic fields up to 7~T using a Quantum Design MPMS-XL magnetometer. The angular-dependent magnetotransport properties were studied in the temperature interval from 2 to 300~K and in magnetic fields up to 9~T employing a Quantum Design PPMS-9 platform equipped with a horizontal rotator. Electrical contacts were made using 50~$\mu$m thick silver wires attached to single-crystalline specimens with silver epoxy paste. The heat capacity was measured over the 2-300~K temperature range by means of relaxation method using the same PPMS system. 

\subsection{Theoretical methods}
The electronic band structure calculations were performed within the density functional theory using the VASP package \cite{VASP1,VASP2,VASP3}. The generalized gradient approximation (GGA) \cite{GGA} and modified Becke-Johnson (MBJGGA) \cite{MBJ} exchange-correlation functionals were employed. Spin-orbit coupling (SOC) was included. The Hubbard $U_\mathrm {eff}$ of 7 eV was assumed for the Tb $4f$ states in the ferromagnetic phase. 
The experimental lattice parameter of the face-centered cubic primitive cell of TbAuPb was used. The plane-wave cut-off and $\bf {k}$-point mesh were set to 600 eV and 16 $\times$ 16 $\times$ 16, respectively.

\section{Experimental results and discussion }
\subsection{Thermodynamic properties}

\begin{figure}
            \centering
            \includegraphics[width=\linewidth]{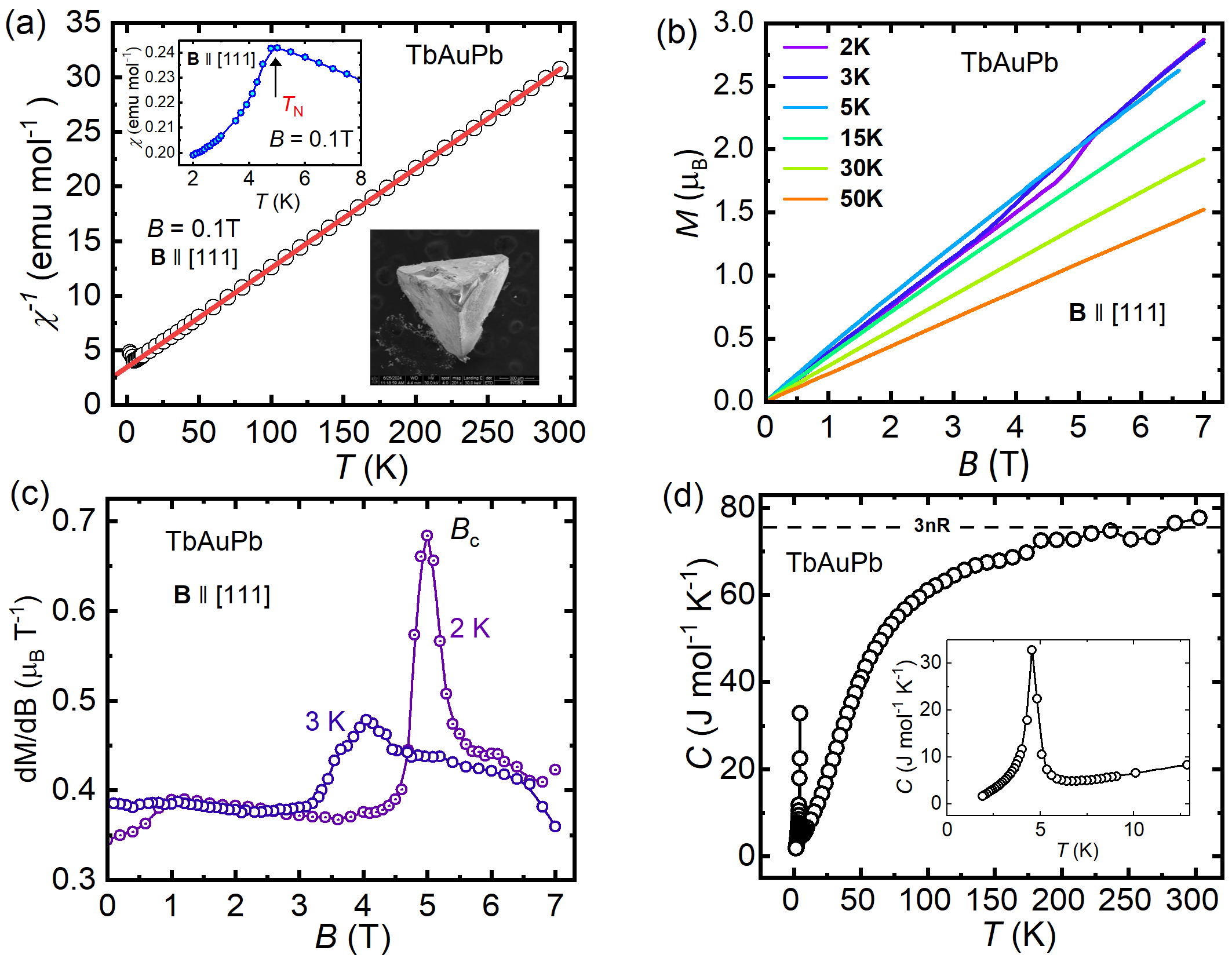}
            \caption{ (a) Temperature dependence of the inverse magnetic susceptibility of single-crystalline TbAuPb measured in a magnetic field 0.1~T applied along the crystallographic [111] axis. The red line represents the Curie-Weiss fit. Upper inset: the low-temperature variation of the magnetic susceptibility measured as in the main panel. The arrow indicates the AFM transition. Lower inset: SEM image of a single crystal of TbAuPb. (b) Magnetic field dependencies of the magnetization in single-crystalline TbAuPb measured at different temperatures along the [111] direction. (c) Field derivative of the magnetization isotherms taken at $T$ = 2 and 3~K. (d) Temperature variation the specific heat of TbAuPb. The horizontal dashed line marks the Dulong-Petit limit. Inset: the low-temperature specific heat data.}
            \label{mag}
        \end{figure}

The temperature variation of the reciprocal magnetic susceptibility $\chi^{-1}(T)$ of TbAuPb measured along the [111] axis is shown in Fig.\ref{mag}(a). Above 50~K, it follows the Curie-Weiss law with the effective magnetic moment $\mu_\mathrm {eff}$ = 9.4 $\mu_\mathrm {B}$ and the paramagnetic Curie temperature $\theta_\mathrm {p}$ = $-38$~K. The experimental value of $\mu_\mathrm {eff}$ is close to the theoretical prediction for a free trivalent Tb ion (9.7 $\mu_\mathrm {B}$). The large negative $\theta_\mathrm {p}$ signals strong antiferromagnetic exchange interactions. Below 50 K, $\chi^{-1} (T)$ slightly deviates from a linear behavior, likely due to crystalline electric field effect. As visualized in the inset to Fig.\ref{mag}(a), the magnetic susceptibility exhibits a peak near $T_\mathrm {N}$ = 5~K, characteristic of the onset of long-range antiferromagnetic (AFM) phase transition.

Fig.\ref{mag}(b) presents the magnetization isotherms $M(B)$ of TbAuPb measured along the [111] direction. In the paramagnetic state, the magnetization varies nearly linearly with $B$. Below $T_\mathrm {N}$, $M$ is also proportional to the magnetic field strength but only up to a critical field $B_\mathrm {c}$ at which a faint inflection occurs in $M(B)$ that possibly indicates a field-induced change in the AFM magnetic structure. As can be inferred from Fig. \ref{mag}(c), $B_\mathrm {c}$ equals about 5~T at $T =$ 2~K and about 4~T at $T =$ 3~K. Above $B_\mathrm {c}$, $M(B)$ retains its quasi-linear behavior.

As displayed in the inset to Fig.~\ref{mag}(d), the second-order AFM transition in TbAuPb manifests itself as a sharp lambda-like singularity in the temperature variation of the specific heat $C(T)$. At higher temperatures, $C(T)$ shows a typical phonon-dominated behavior (see Fig.~\ref{mag}(d)) and tends towards saturation at a value of 75\;$\rm{J\;mol^{-1}\;K^{-1}}$, which corresponds to the Dulong-Petit limit 3$nR$, where $n$ is the number of atoms in  formula unit and $R$ stands for the universal gas constant.

\subsection{Magnetotransport properties}
  
\begin{figure*}
        \centering
        \includegraphics[width=\linewidth]{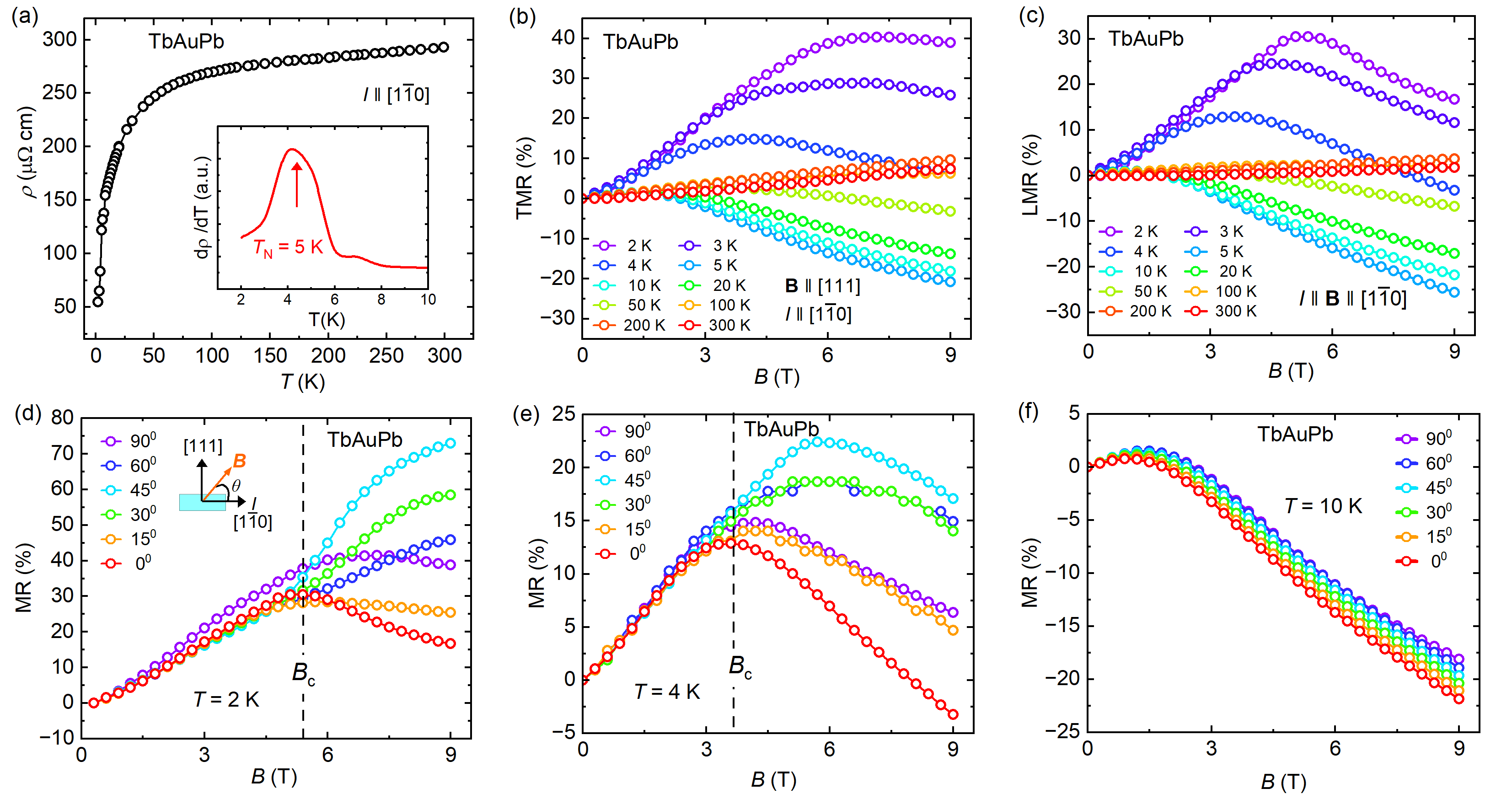}
        \caption{(a) Temperature variation of the electrical resistivity of single-crystalline TbAuPb measured with electric current flowing along the [1$\bar 1$0] direction. Inset: temperature derivative of the resistivity measured at low temperatures. (b) Transverse magnetoresistance isotherms of TbAuPb taken at various temperatures with electric current flowing along the [1$\bar 1$0] direction and magnetic field applied along the [111] axis. (c) Longitudinal magnetoresistance isotherms of TbAuPb taken at various temperatures with electric current flowing along the [1$\bar 1$0] direction and magnetic field applied parallel to the current. (d-f) Magnetoresistance of TbAuPb measured at 2~K, 4~K, and 10~K, respectively,  with electric current flowing along the [1$\bar 1$0] direction in magnetic field applied at different angles with respect to the current direction, as shown in the inset to panel (d).}
        \label{res}
\end{figure*}

The temperature dependence of the electrical resistivity $\rho(T)$ of TbAuPb measured with electric current $\textbf{I}$ flowing along the [1$\bar 1$0] direction is presented in Fig.~\ref{res}(a). The compound exhibits a semimetallic-like behavior, similar to that found for GdAuPb \cite{Liu2024}.The value of the residual resistivity ratio (RRR = $\rho(300~{\rm K})$/$\rho(2~{\rm K})$) is approximately 5.3. At $T_\mathrm{N}$ = 5~K, the $\rho(T)$ curve shows a slope change that results in a peak in its temperature derivative (see the inset to Fig.~\ref{res}(a)). This feature is due to rapid reduction in spin-disorder scattering of charge carriers in the AFM state.

The magnetoresistance (MR = $\frac{\rho(B)-\rho(B = 0)}{\rho(B = 0)}\times100\%)$ of single-crystalline TbAuPb was measured as a function of the external magnetic field magnitude in both transverse (TMR; $\textbf{B} \perp \textbf{I}$) and longitudinal (LMR; $\textbf{B} \parallel \textbf{I}$) configuration with electric current $\textbf{I} \parallel [1 \bar{1} 0]$. As can be inferred from Fig.~\ref{res}(b), TMR is positive in the ordered state. However, in strong magnetic fields, a negative contribution becomes dominant (probably originating from de Gennes - Friedel mechanism), which gives rise to the formation of a broad maximum near the critical field $B_\mathrm{c}$. With increasing temperature, the magnitude of TMR decreases and the maximum shifts to lower values of the field. The TMR isotherms taken above $T_\mathrm{N}$ show a faint positive maximum near 1.5 T (hardly visible in Fig.~\ref{res}(b)) but in stronger fields TMR is negative, and its absolute magnitude decreases with increasing temperature. This behavior can be explained by the field-induced alignment of magnetic moments in the paramagnetic state \cite{Gennes}, which becomes less effective at higher temperatures. At temperatures $T \geq$ 100~K, TMR shows slightly positive values, probably resulting predominantly from the Lorentz effect. It is noteworthy that the behaviour of TMR in TbAuPb differs distinctly from that observed in GdAuPb, in which a substantial, unsaturated positive magnetoresistance was detected at low temperatures \cite{Liu2024}.  

Fig.~\ref{res}(c) presents the LMR data of TbAuPb. Below $T_\mathrm{N}$, LMR behaves very similarly to TMR, forming a broad maximum around $B_\mathrm{c}$ due to the interplay of positive and negative contributions. However, the latter is stronger than in the case of TMR, which leads to a faster decrease in the LMR magnitude in high fields, and even to a change in the sign of the isotherm measured at 5 K. In the paramagnetic region, LMR mimics the behavior of TMR.

MR was also measured at different angles $\theta$ between $B$ and $I$, as shown in the inset of Fig.~\ref{res}(d). In the AFM state ($T <$ 5~K), the MR curves taken at different $\theta$ overlap below $B_\mathrm{c}$, and diverge in stronger fields. Interestingly, in the latter region, the negative component to MR measured at $T =$ 2~K almost disappears for angles close to 45$\degree$, and positive MR attains a value as large as 75\% in $B =$ 9~T. For MR taken at $T =$ 4~K (see Fig.~\ref{res}(e)), the effect is less distinct, and weaker negative component brings about only shift of the maximum in MR to fields significantly stronger than $B_\mathrm{c}$. The observed behavior suggests that in high magnetic fields the Tb magnetic moments are arranged in a non-collinear manner. However, a neutron diffraction experiment is required to verify this conjecture. Eventually, as visualized in Fig.~\ref{res}(f), in the paramagnetic region, MR is mostly negative (except for weak magnetic fields) and its absolute magnitude increases gradually when transitioning from the transverse to the parallel configuration of $\textbf{B}$ and $\textbf{I}$, however the overall change is rather small. 

\begin{figure*}
        \centering
        \includegraphics[width=0.8\linewidth]{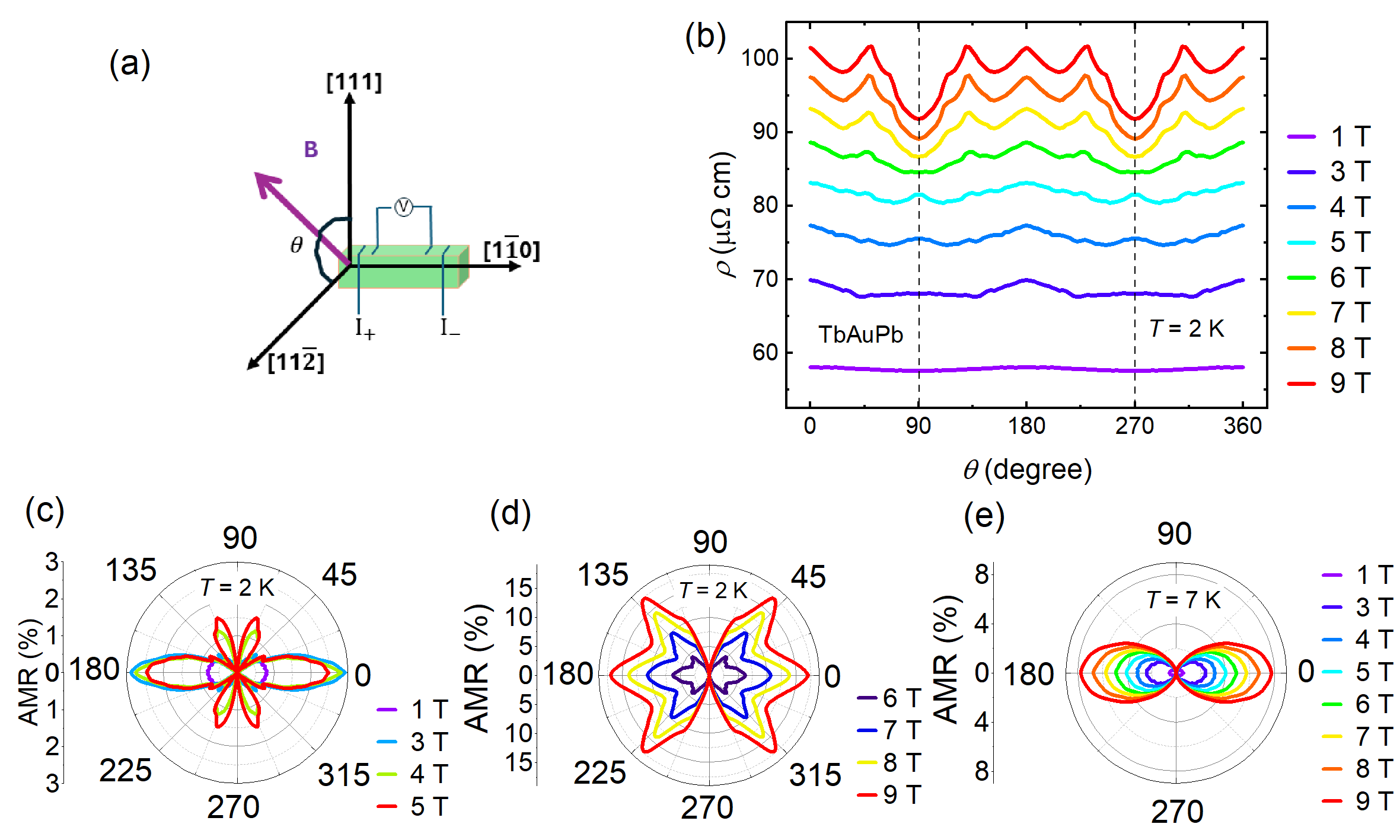}
        \caption{(a) Mutual arrangement of magnetic field and electric current in measurements of the angular magnetoresistance in single-crystalline TbAuPb performed in transverse configuration. (b) Angular variations of the electrical resistivity of TbAuPb measured at 2~K as shown in panel (a) in different applied magnetic fields. For clarity, the subsequent curves were shifted upwards by a multiple of 5~$\mu\ohm$ cm. (c) Polar plots of the angular magnetoresistance of TbAuPb measured as shown in panel (a) in the AFM state below the spin-reorientation transition. (d) Polar plots of the angular magnetoresistance of TbAuPb measured as shown in panel (a) in the AFM state above the spin-reorientation transition. (e) Polar plots of the angular magnetoresistance of TbAuPb measured as shown in panel (a) in the paramagnetic state in different magnetic fields.}
        \label{amr}
    \end{figure*}

Measurements of the electrical resistivity of TbAuPb in the transverse configuration ($\textbf{B} \perp \textbf{I}$) were extended by examining its angular dependence $\rho({\theta})$ at $T =$ 2~K, when the magnetic field was rotated by 360$\degree$ in the plane (1$\bar{1}0$), while the electric current flowed along the direction [1$\bar{1}$0] (cf. Fig.~\ref{amr}(a)). As visualized in Fig.~\ref{amr}(b), in fields $\leq{5~T}$, $\rho({\theta})$ forms peaks centered at 90$\degree$ and 270$\degree$, which invert into valleys in stronger fields. This peak-valley inversion corresponds to the spin-reorientation transition observed at $B_\mathrm {c}$ = 5~T. A similar inversion in $\rho({\theta})$ was reported for the Weyl semimetal candidate TbPtBi in both AFM and paramagnetic states, and attributed to the magnetic field-induced changes in the electronic band structure \cite{Chen2023}. 

In order to identify a possible topological phase transition in TbAuPb, 
the anisotropic magnetoresistance (AMR = $ \frac{\rho(\theta)-\rho(\theta = 90\rm{\degree})}{\rho(\theta = 90\rm{\degree})}\times100 \%$) was determined from the $\rho(\theta)$ data taken below and above $T_\mathrm {N}$ in magnetic fields ranging from 1~T to 9~T, applied as shown in Fig.~\ref{amr}(a). At $T =$ 2~K, AMR exhibits a two-fold symmetry up to 3~T with the peak centered at 0$\degree$ rising in magnitude with increasing field (see Fig.~\ref{amr}(c)). In stronger fields, additional peaks emerge, centered around 68$\degree$, 123$\degree$, 203$\degree$, and 293$\degree$ indicating that higher-order symmetry terms become significant in AMR on approaching the spin-reorientation at $B_\mathrm {c}$. With increasing field, the peaks centered at  0$\degree$ and 68$\degree$ merge, and an unusual butterfly-shaped AMR pattern forms with peaks centered around 0$\degree$, 45$\degree$, 135$\degree$, 180$\degree$, 225$\degree$, and 315$\degree$ (see Fig.~\ref{amr}(d)). In contrast to this fairly complex behavior of AMR in the AFM state, AMR measured above $T_\mathrm {N}$ exhibits a conventional two-fold symmetry, independent of the magnetic field strength, with the magnitude systematically increases with increasing field (see Fig.~\ref{amr}(e)).

\begin{figure*}
        \centering
        \includegraphics[width=\linewidth]{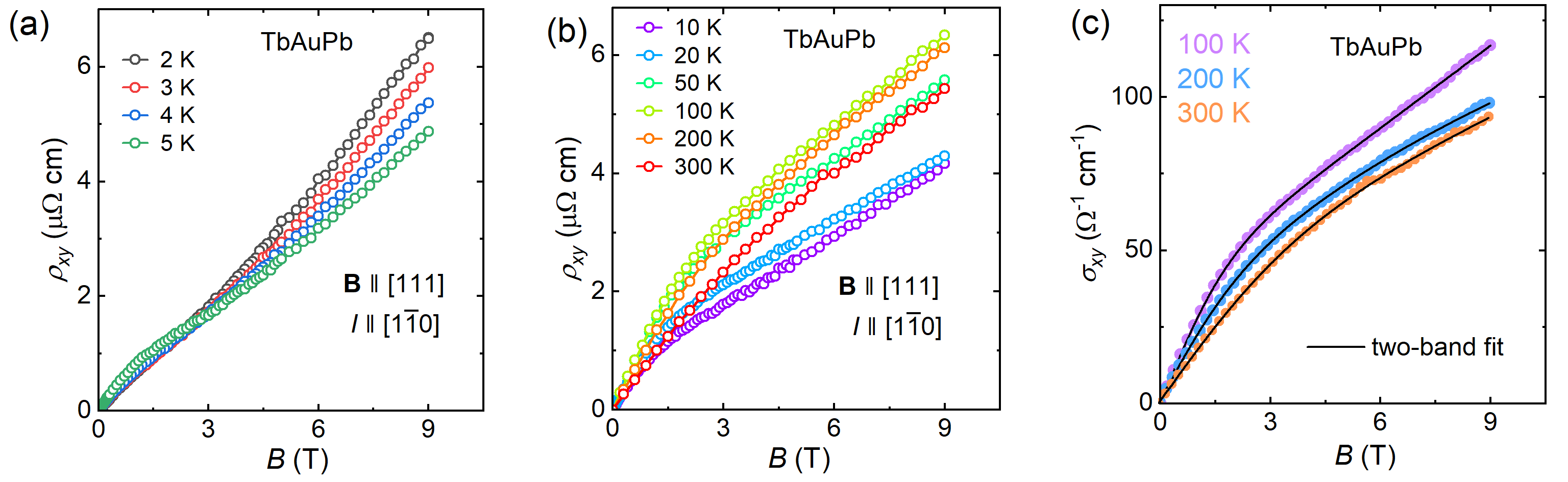}
        \caption{(a) Hall resistivity of single-crystalline TbAuPb measured as a function of magnetic field at different temperatures in the AFM state with electric current flowing along the [1$\bar 1$0] direction and magnetic field applied along the [111] axis. (b) Hall resistivity of TbAuPb measured at different temperatures as described in panel (a) in the paramagnetic state. (c) The Hall conductivity at high temperatures $T >$ 100~K. The solid black lines represent the two-band model fits (see Eq.~\ref{twobandmodel})}.
        \label{Hall}
\end{figure*}

The Hall resistivity $\rho_{xy}$ of TbAuPb was measured at different temperatures as a function of magnetic field with electric current flowing along the [1$\bar 1$0] direction and magnetic field applied along the [111] axis, As shown in Figs.~\ref{Hall}(a) and (b), $\rho_{xy}(B)$ is positive, which suggests that holes are the majority charge carriers. 
At $T_\mathrm{N}$, $\rho_{xy}$ exhibits a pronounced bump near 1.5~T. Below $T_\mathrm{N}$, a clear change in the slope of $\rho_{xy}(B)$ occurs near $B_\mathrm {c}$, which likely originates either from a modification of the anomalous Hall effect in the AFM state or from the reconstruction of the Fermi surface associated with the AFM state \cite{Kurumaji2024, Kusakabe2025}. It is worth noting that the large anomalous Hall was reported for GdAuPb in the AFM state \cite{Liu2024}. However, in the present case, the anomalous Hall resistivity can not be reliably separated from the experimental Hall data because of multiband contributions, and also the magnetization does not saturate within the measured field range. Therefore, measurements at higher magnetic fields are required.
In order to quantify multiband contributions, the Hall conductivity, in TbAuPb was analyzed in terms of a two-band model by fitting the experimental data collected at $T \geq$ 100~K (at lower temperatures, TMR is negative preventing reliable application of this approach) with the formula

\begin{equation}
    \sigma_{xy} = eB\sum_i^m\left[\frac{n_i\mu_i^2}{1+(\mu_iB)^2}\right] ,
    \label{twobandmodel}
\end{equation}

\noindent where $m$ is the number of Fermi pocket, $n_{i}$ and $\mu_{i}$ represent the carrier concentration and mobility of charge carriers from $i^{th}$ Fermi pocket, respectively. The results are displayed in Fig.~\ref{Hall}(c) and summarized in Table 1.

\begin{table}
    \caption{Concentrations and mobilities of the charge carriers in TbAuPb at high temperatures}
    \begin{tabular}{ccccc}
        \hline
        T  & $n_1$   & $\mu_1$  & $n_2$   & $\mu_2$ \\
        (K) & ($\rm{cm^{-3}}$) & ($\rm{cm^{2}V^{-1}s^{-1}}$) & ($\rm{cm^{-3}}$) & ($\rm{cm^{2}V^{-1}s^{-1}}$)\\
        \hline 
        100   & 1.5 $\times 10^{19}$  &   218 & 8.6$\times 10^{16}$ &  3366 \\ 
        200  &  3.8 $\times 10^{18}$   &  412 & 9.9 $\times 10^{16}$ & 2695\\ 
        300  &  6.5 $\times 10^{19}$   &  297 & 2.0 $\times 10^{17}$ & 1576\\ 
        \hline
    \end{tabular}
    \label{tab:fitting_parameters}
\end{table}

\section{Electronic structure}
\label{DFT}

\begin{figure}[]
    \centering
    \includegraphics[width=\linewidth]{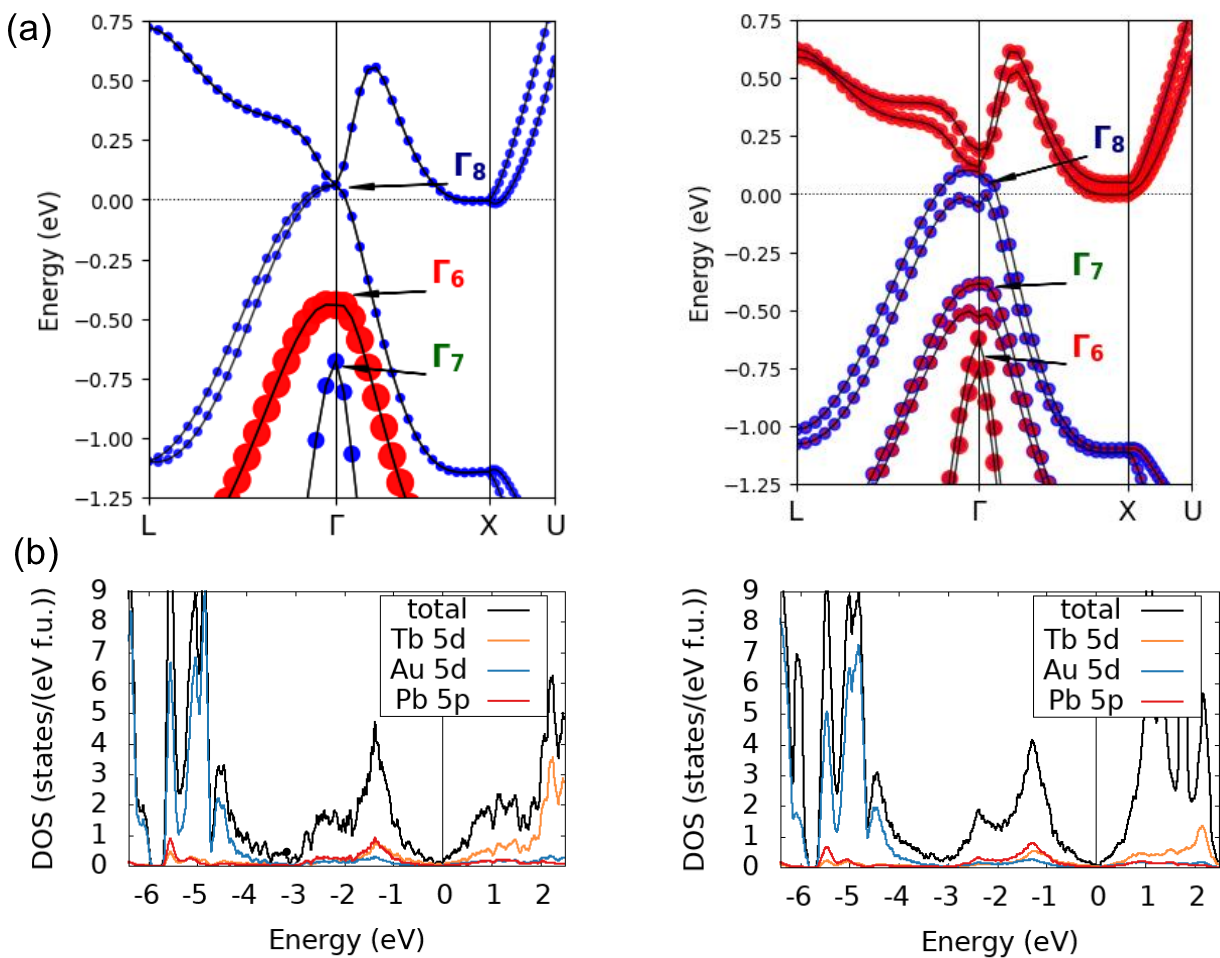}
    \caption{(a) Electronic structure of nonmagnetic TbAuPb calculated using the MBJGGA potential (left) and ferromagnetic  phase with $M \parallel [001]$ of TbAuPb calculated within the MBJGGA+$U$ approach (right). Bands bearing $p$- and $s$-type character are marked with blue and red dots, respectively. (b) Total and partial density of states in TbAuPb calculated as specified in panel (a).}
    \label{fig:BAND}
\end{figure}

The electronic band structure calculations performed within the MBJGGA approach indicated that nonmagnetic TbAuPb would be a semimetal (see Fig.~\ref{fig:BAND}(a)) with small Fermi surface (FS) pockets located at the $\Gamma$ and $X$ points in the Brillouin zone (BZ). The similar FS pockets have also been formed in semimetallic $R$PtBi \cite{Suzuki2016} and in recently explored $R$AuSn family \cite{Ueda2025}. 

In TbAuPb, the electronic bands crossing the Fermi level ($E_{\rm F}$) exhibit a $p$-type character, whereas the $s$-type contributions are located well below $E_{\rm F}$. Such an inversion of a typical band order in face-centered cubic systems may be connected with the formation of topologically nontrivial states, as intensively investigated in recent years for various HH materials \cite{Al-Sawai2010, Feng2010}. It is worth noting that strong spin-orbit coupling (SOC) in TbAuPb leads to a pronounced splitting of the valence bands along the $L$-$\Gamma$ line, however the bands along the $\Gamma$-$X$ line remain degenerate. In topologically non-trivial materials, band-inversion leads to robust surface states with an odd number of Fermi-level crossings. These states have been extensively studied theoretically and experimentally using angle-resolved photoemission spectroscopy (ARPES) \cite{Chen2009, Neupane2012}. However, in HHs, particularly single crystalline $R$PtBi, ARPES studies have revealed trivial metallic surface states that cross $E_{\rm F}$ an even number of times, resulting in zero Berry curvature \cite{Liu2011}. This observation is in disagreement with the theoretical band structures for bulk systems that predict an inverted band-gap \cite{Feng2010}. The classification of topological phases of HH materials is thus still under debate.

Our experiments have shown that the ground state in TbAuPb is antiferromagnetic, and it would be appropriate to perform calculations of the electronic structure assuming an AFM phase. Unfortunately, without knowledge of the magnetic structure of this material, the results of such calculations may prove questionable or completely inadequate. It is certainly worthwhile to perform them once the configuration of magnetic moments in the low- and high-field AFM phases of TbAuPb has been determined using neutron diffraction. On the other hand, it seems appropriate to examine at the band structure of TbAuPb in a ferromagnetic (FM) state, which can be induced in this material by applying a very strong magnetic field.  Therefore, shown in Fig.~\ref{fig:BAND} are the results of our calculations performed using a MBJGGA+$U$ approach to the FM phase, with moments aligned along the [001] direction in TbAuPb. As can be inferred from this figure, the electronic bands have similar shapes to those in the nonmagnetic phase, however, their degeneracy is lifted \and partially cross the Fermi level in the vicinity of the $X$ point. The band character of conduction bands is clearly $s$-type, while the remaining FS surface sheets are dominated by some $p$-type contributions, corresponding to a topologically trivial band order. On the other hand, the $\Gamma_7$ and $\Gamma_6$ bands exhibit the $s$-type character to some extent, which indicates some transient state between topologically trivial and nontrivial band order. 

Fig.~\ref{fig:BAND}(b) presents the density of states (DOS) in TbAuPb determined for the nonmagnetic and FM phases, respectively. Regardless  of the magnetic phase, $E_{\rm F}$ is located in a pseudogap, which highlights the semimetallic character of the compound. In the nonmagnetic phase, DOS at $E_{\rm F}$ amounts to 0.44 states eV$^{-1}$ f.u.$^{-1}$, which correspond to the Sommerfeld coefficient value of 1.05 mJ\,mol$^{-1}$K$^{-2}$. Despite the volume of FS pockets being relatively smaller, the number of bands crossing $E_{\rm F}$ in the nonmagnetic phase is larger compared to the FM case. The valence band region in the vicinity of $E_{\rm F}$ is dominated by the Pb $5p$ and Tb $5d$ states, whereas the $5d$ contributions from Au ions are located about 4 eV below $E_{\rm F}$. In turn, the conduction region is dominated by the unoccupied Tb $5d$ and Pb $5p$ states.

In $RT$Bi ($T$ = Pt or Pd) family, the observed anomalies in magnetotransport experiments like negative longitudinal magnetoresistance and anomalous Hall effect  are mainly linked to two phenomena: (i) the formation of Weyl nodes \cite{Hirschberger2016, Guo2018, Shekhar2018} and (ii) the large Berry curvature due to the avoided band-crossings \cite{Suzuki2016, Zhang2020, Zhu2020}. Strong SOC in HHs opens the gap at the nodal points and these avoided band-crossings form the Berry curvature hot-spots, which results in large anomalous Hall effect \cite{Zhu2023}. In TbAuPb, Weyl nodes were not observed near the Fermi level and there is no signature of chiral anomaly in the magnetotransport as negative magnetoresistance is small and isotropic.

\begin{figure}[]
    \centering
    \includegraphics[width=\linewidth]{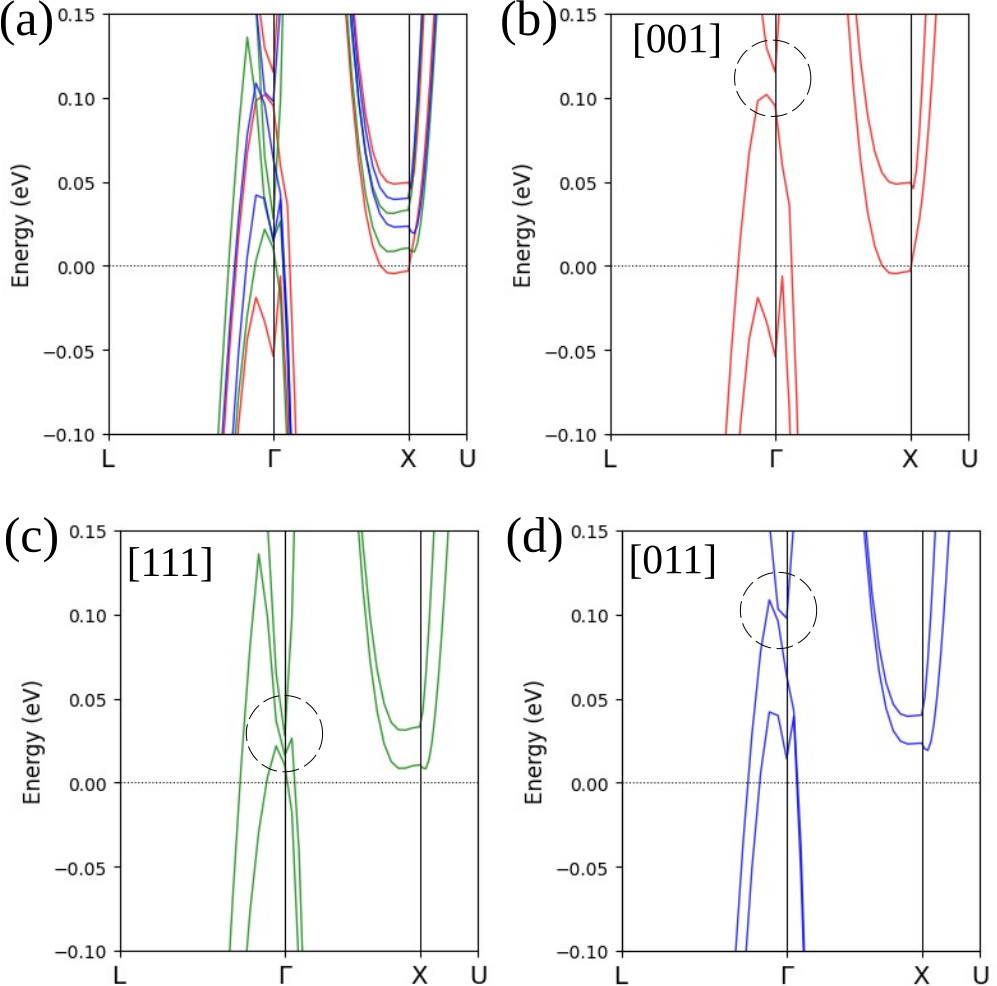}
    \caption{Band structures calculated (MBJGGA+$U$) for the ferromagnetic phase of TbAuPb with magnetization directions (a) [001] (red), [111] (green), [011] (blue), (b) [001], (c) [111], and (d) [011]. The band anti-crossings are marked with dashed circles.}
    \label{fig:FMBAND}
\end{figure}

The electronic band structure in HH materials is sensitive to the direction of magnetization and can be strongly affected by the spin textures, as  discussed recently for $R$AuSn family of compounds \cite{Ueda2025}. In the present case, the electronic band structure in the FM state is obtained by considering the alignment of magnetic moments in various high symmetry directions, as shown in Fig.~\ref{fig:FMBAND}. The avoided band-crossings (dashed circle) have been observed in all cases considered here, we can thus predict that, similarly to those in other HHs, these anti band-crossings may contribute to the observed anomalous magnetotransport properties in TbAuPb. However, more detailed Berry curvature and theoretical calculation of anomalous Hall conductivity is needed to prove this hypothesis.

In Fig.S2, the electronic band-structure of non-magnetic TbAuPb was presented for decreased (pressure) and increased (tensile stress) unit cell volumes with respect to the equilibrium one in order do analyze the possible effect of these conditions on the electronic structure and thereby affect the transport properties. We find that the position of $E_{\rm F}$ does not change significantly under pressure, and the inverted band-order remains robust. However, the position and shape of the $\Gamma_6$ band evolve systematically with increasing  pressure, and the system may become a trivial semimetal under high pressure. TbAuPb appears to be close to a topological critical point, where external factors such as magnetic field or hydrostatic pressure can change the band topology from non-trivial to trivial, as also shown for GdAuPb \cite{Liu2024, Saeidi2018}.

\section{Summary}

The single crystals of TbAuPb were grown using the self-flux method. They were found to crystallize with a cubic structure of the MgAgAs-type, thus being a representative of the half-Heusler family. Magnetic and specific heat measurements established the AFM ordering below $T_\mathrm{N}$ = 5~K. The low-field magnetic structure transforms into a high-field AFM phase at the critical field $B_\mathrm{c}$. At 2 K, $B_\mathrm{c} =$ 5~T, and decreases with increasing temperature. The electronic transport reveals semi-metallic nature of the compound. In the ordered state, both transverse and longitudinal magnetoresistance are positive, yet a negative contribution becomes significant above $B_\mathrm{c}$. In the paramagnetic state, negative magnetoresistance was observed in both transverse and longitudinal configurations. The Hall resistivity indicates the presence of anomalous Hall effect in the AFM state together with multiband contribution  with holes being the charge carriers. The angular magnetoresistance measured below $T_\mathrm{N}$ exhibits an anomalous butterfly-like pattern above 5~T. In contrast, in the paramagnetic state, it exhibits a conventional two-fold symmetry. Our first-principles calculations supported the experimental finding of the semi-metallic character of TbAuPb with the formation of small Fermi pockets. Due to strong spin-orbit interaction, a band inversion occurs in the nonmagnetic phase that becomes trivial in the ferromagnetic state. In order to perform calculations for the AFM phase, the determination of the actual magnetic structures is required by means of neutron diffraction. In this study, we establish TbAuPb, a new half-Heusler compound, as a potential platform to study the multiple phenomena like metamagnetic transition, anomalous Hall effect, multiband transport, and non-trivial topology, simultaneously. Further high magnetic-field measurements and more detailed first-principles calculations are required to extract the anomalous Hall effect and explain its origin.

\section{Acknowledgments}
This work was supported by National Science Centre (Poland) under project No. 2021/40/Q/ST5/00066. The DFT calculations were performed at Wroclaw Center for Networking and Supercomputing (Project No. 158).

\bibliography{mybib.bib}

\clearpage
\newpage
\onecolumngrid

\begin{center}
  \textbf{\Large Supplementary Material}\\[.2cm]
  \textbf{\large Thermodynamic and electrical transport properties of the half-Heusler plumbide TbAuPb }\\[.2cm]
  A.Agarwal, S. Chatterjee, Maciej J. Winiarski, O. Pavlosiuk, D. A. Kowalska, P. Wi\'{s}niewski, and D. Kaczorowski\\[.2cm]
  {\itshape
  	\mbox{Institute of Low Temperature and Structure Research, Polish Academy of Sciences, Okólna 2, 50-422 Wrocław, Poland}\\ [.2cm]
	}

\end{center}

\setcounter{equation}{0}
\renewcommand{\theequation}{S\arabic{equation}}
\setcounter{figure}{0}
\renewcommand{\thefigure}{S\arabic{figure}}
\setcounter{section}{0}
\renewcommand{\thesection}{S\arabic{section}}
\setcounter{table}{0}
\renewcommand{\thetable}{S\arabic{table}}
\setcounter{page}{1}


\section{EDX analysis}
Fig.~S1 shows an example of the energy-dispersive X-ray (EDX) spectra collected on the single crystals of TbAuPb, which were grown in this work. Analysis of the EDX data yielded the Tb:Au:Pb ratio of 1:0.97:1.05 which is close to the ideal equiatomic composition. Small peaks caused by the presence of carbon and oxygen deposited on the specimen surface were observed in the EDX spectra. Prior to physical transport measurements, all samples were carefully polished to remove any residual contaminants.
 
\begin{figure}[h!]
    \centering
    \includegraphics[width=0.5\linewidth]{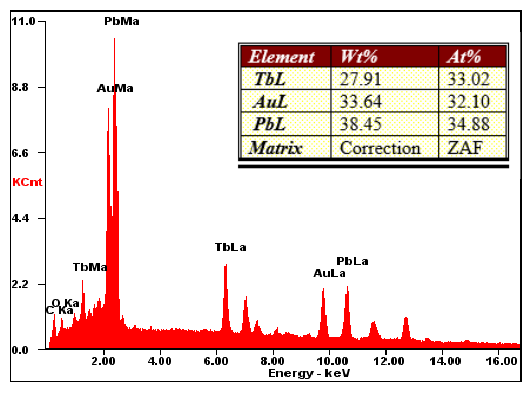}
    \caption{EDX spectrum of TbAuPb and the chemical composition of the crystal examined.}
    \label{fig:edx}
\end{figure}

\section{single-crystal X-ray diffraction}
Single-crystal X-ray diffraction data were collected on an Oxford Diffraction X'Calibur diffractometer with a CCD Atlas detector using Mo K$\alpha$ radiation ($\lambda$ = 0.71073 \AA). Data collection and reduction were performed with CrysAlisPro 1.171.42.93a (Rigaku Oxford Diffraction, 2023). A numerical absorption correction was applied using Gaussian integration over a multifaceted crystal model. The crystal structure was solved by direct methods and refined by full-matrix least-squares on $F^2$ using SHELX (Sheldrick, 2015) via Olex2 (Dolomanov et al., 2009). Crystallographic and refinement details are given in Table \ref{tab:xrd1}, and the data were deposited with the Cambridge Crystallographic Data Centre (deposition number CSD 2522510).

\begin{table} [h]
    \caption{Single-crystal X-ray diffraction experimental results}
    \begin{tabular}{p{0.45\linewidth} p{0.3\linewidth}}
        \hline
        \\
        $\textbf{Crystal Data}$  &    \\
        
        Chemical formula  &  TbAuPb\\
        Formula weight (g mol$^{-1}$)   &  563.08\\
        Crystal System, space group    &   Cubic, $F\bar{4}3m$ (no. 216)\\
        Temperature (K) & 295\\
        Unit cell parameters (\AA) & $a$ = 6.745(4)\\
        Unit cell volume (\AA$^3$) & 306.9(5)\\
        Z / calculated density (g cm$^{-3}$) & 4 / 12.188\\
        $\mu$ (mm$^{-1}$) & 124.88\\
        Crystal size (mm) & 0.18$\times$0.13$\times$0.07\\
        &\\
        $\textbf{Data Collection}$  &    \\
        Theta range &  5.2$\degree$--27.3$\degree$\\
        Reflections measured/ unique/ observed [$I$ $>$ 2$\sigma(I)$] & 320/ 56/ 56\\
        R$_{int}$ & 0.044\\
        (sin $\theta$/$\lambda_{max}$) \AA$^{-1}$ & 0.646\\
        Limitig indices & $h$ = -5 $\rightarrow$ 8,  $k$ = -4 $\rightarrow$ 8,  $l$ = -7 $\rightarrow$ 7\\
        \\
        $\textbf{Refinement}$  &    \\
        $R$ indices [$F^2 > 2\sigma(F^2)$] &  $R_1$ = 0.019, $wR_{2}$ = 0.048\\
        Goodness of fit on $F^2$ & 1.03\\
        Data/ parameters/ restraints & 56/ 5/ 0 \\
        Extinction coefficient & 0.0041(6)\\
        Largest difference peak and hole (e \AA$^{-3}$) & 1.37, -0.87\\
        Absolute structure parameter & -0.06(3)\\
        \hline
        
    \end{tabular}
    \label{tab:xrd1}
\end{table}

\section{Electronic band structure}
The electronic band structure of TbAuPb at different pressures in the nonmagnetic state. 

\begin{figure}[h!]
    \centering
    \includegraphics[width=\linewidth]{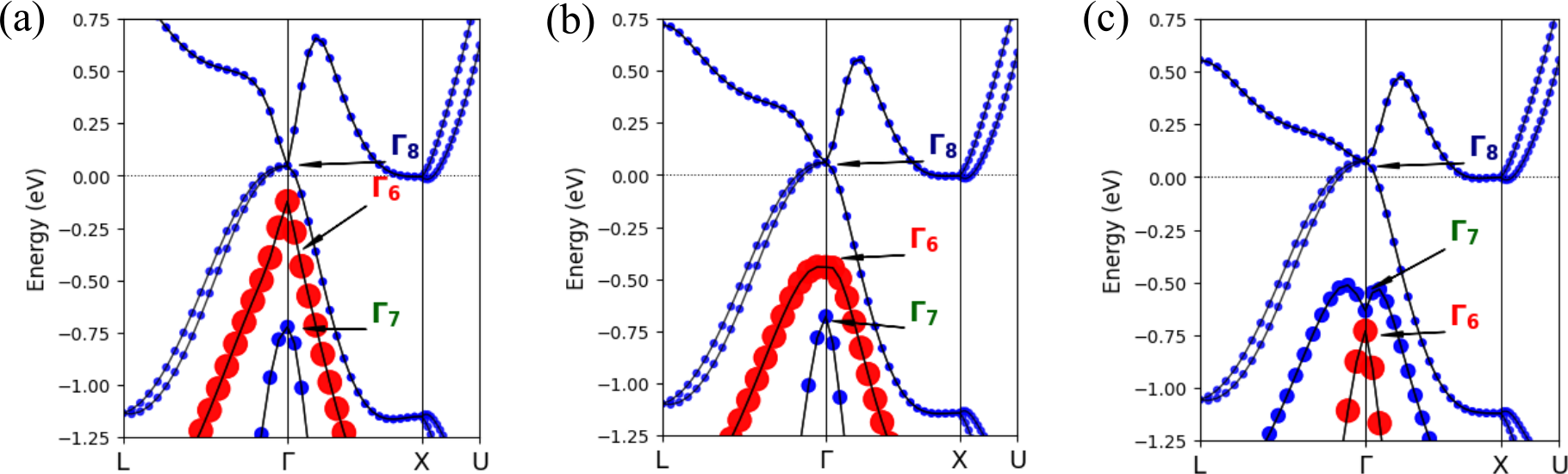}
    \caption{Band structures calculated using the MBJGGA potential for nonmagnetic TbAuPb with the lattice parameter {\it a} of (a) -1.65 \%, (b) 0 \%, and (c) +1.65 \% with respect to the equilibrium one ({\it $a_0$}). The $p$- and $s$-type band characters are marked with blue and red dots, respectively.} 
    \label{pressure}
\end{figure}

\end{document}